\documentclass[aps,preprint,showpacs,superscriptaddress]{revtex4}
    \usepackage{amssymb}
    \usepackage{graphicx}
    \usepackage{amsfonts}              
    \usepackage{amssymb,amsmath}
    \usepackage{bm}
    \usepackage{comment}

\begin{document}
   \title[Rashba spin-orbit coupling of Landau levels]{Local spin polarization of Landau levels under Rashba spin-orbit coupling}   
      \author{T. Fujita}       
    \affiliation{Information Storage Materials Laboratory, Electrical and Computer Engineering Department, National University of Singapore, 4 Engineering Drive 3, Singapore 117576}
    \affiliation{Data Storage Institute, DSI Building, 5 Engineering Drive 1, (off Kent Ridge Crescent, National University of Singapore) Singapore 117608}
    \affiliation{Computational Nanoelectronics and Nano-device Laboratory, Electrical and Computer Engineering Department,
National University of Singapore, 4 Engineering Drive 3, Singapore 117576}
    \author{M. B. A. Jalil}
    \affiliation{Information Storage Materials Laboratory, Electrical and Computer Engineering Department, National University of Singapore, 4 Engineering Drive 3, Singapore 117576}
    \affiliation{Computational Nanoelectronics and Nano-device Laboratory, Electrical and Computer Engineering Department,
National University of Singapore, 4 Engineering Drive 3, Singapore 117576}
    \author{S. G. Tan}
    \affiliation{Data Storage Institute, DSI Building, 5 Engineering Drive 1, (off Kent Ridge Crescent, National University of Singapore) Singapore
    117608}
    \affiliation{Computational Nanoelectronics and Nano-device Laboratory, Electrical and Computer Engineering Department,
National University of Singapore, 4 Engineering Drive 3, Singapore 117576}
    \author{F. Wan}
        \affiliation{Information Storage Materials Laboratory, Electrical and Computer Engineering Department, National University of Singapore, 4 Engineering Drive 3, Singapore 117576}
    
\begin{abstract}
We investigate the local spin polarization texture of Landau levels under Rashba spin-orbit coupling in bulk two-dimensional electron gas (2DEG) systems. In order to analyze the spin polarization as a function of two-dimensional coordinates within the 2DEG, we first solve the system eigenstates in the symmetric gauge. Our exact analytical wavefunction solutions are shown to be gauge invariant with solutions obtained in the commonly used Landau gauge. We illustrate the two-dimensional spatial spin profile for a single Landau level and suggest means to measure and utilize the local polarization in practice.
\end{abstract}
 

\maketitle

\section{Introduction}
Spin dependent transport phenomena in low dimensional systems have
attracted considerable attention in recent years because of their
potential application in information processing and storage
devices \cite{sa,zutic,awsch}. A paradigmatic proposal is the spin
field-effect-transistor which utilizes the
gate-controllable \cite{nitta} Rashba spin-orbit coupling
(SOC) \cite{rashba1,rashba2} in two-dimensional electron gases
(2DEGs) to control the spin rotation of electrons as they propagate
across the device \cite{datta,mireles,fujita}. The Rashba SOC results
from the structural inversion asymmetry of the microscopic
confinement potential formed at the interface of semiconductor
heterostructures \cite{rashba1,rashba2}. There is much interest
recently in 2DEG systems with SOC and external magnetic fields. By
applying a perpendicular magnetic field to the 2DEG system, the SOC
competes with Zeeman spin-splitting and this interplay leads to
further modification of band structure and other interesting
results. A few examples are resonant spin-Hall conductance \cite{shen,shun} due to induced degeneracies of Landau levels at
certain values of magnetic field \cite{zhigang}, modified
magneto-optical transition spectrums \cite{yang}, beating patterns
in the density of states and longitudinal resistivity \cite{wang},
and altered Hall conductance \cite{demikhovskii} which differs from
the quantized values in the integer quantum Hall (IQH) regime. We
note that previous works (including all of the above) perform their
analyses in the Landau gauge. While the use of the Landau gauge is
perfectly valid due to gauge invariance, the form of the
wavefunction in this gauge does not capture the natural, rotational
symmetry of the eigenstates. For example, in the presence of Rashba
SOC, the spin polarization of eigenstates exhibits interesting
spatial textures whose features cannot adequately be reflected by
the wavefunctions obtained in the Landau gauge. The study of the locally varying spin polarization within a 2DEG may have a number of
interesting applications. For instance, a well-controlled spin
texture with distinct spatial modulation may be used as a resolution
test for surface spin probe techniques. Additionally, it may be
possible, by means of some localized probes, to harness an efficient
spin current source from spatial regions with high spin
polarization. Here, the spatial separation of spins is reminiscent of the optical dispersion (spatial separation of optical frequencies) found in monochromators, suggesting that it may be used as a form of spin filter.
\\In this article, we theoretically study the local spin polarization of Landau levels in the presence of Rashba SOC within an infinite 2DEG. To do so, we first present analytical solutions of the eigenstates of the system in the rotationally isotropic symmetric gauge. We demonstrate the
gauge equivalence of our solutions with previously known solutions
obtained in the Landau gauge. Finally, we show the spatial distribution of the spin components of
Landau level states in the presence of Zeeman and Rashba SOC
effects.
\section{Theory}
\subsection{Landau levels in the symmetric gauge}
We first solve the Landau level wavefunctions without the Rashba SOC
and Zeeman interactions in the symmetric gauge i.e.\ the IQH states.
The wavefunctions form an orthonormal set which we use as the basis
functions in solving the complete system. Under an external vertical
magnetic field $\mathbf{B}$, the Hamiltonian of a spinless and
otherwise free electron in a 2DEG is written as $\mathcal{H}_0 =
\mathbf{\Pi}^2/2m_e$, where $\mathbf{\Pi} = \mathbf{p}+e \mathbf{A}$ is
the covariant momentum under the vector potential $\mathbf{A}$ which
satisfies $\text{curl }\mathbf{A}=\mathbf{B}$, $-e$ is the electron charge, and $m_e$ the effective electron mass. Fixing the magnetic field $\mathbf{B}$ does not uniquely
determine the vector potential, i.e.\ there is a gauge freedom. For
a magnetic field that is perpendicular to the plane of the 2DEG,
pointing in the $\hat{z}$-direction by convention, the Landau (L)
gauge is given by $\mathbf{A}^{L} = (-B_zy,0,0)$ whilst the symmetric
(sym) gauge is given by  $\mathbf{A}^{\text{sym}} = B_z/2(-y,+x,0)$,
where $B_z$ is the magnetic flux density (in Tesla) of the external
field and $x,y$ are spatial coordinates in the plane of the 2DEG. In
the presence of a uniform magnetic field, the system exhibits both
translational and rotational symmetry about the $\hat{z}$-axis.
Under the Landau gauge, it is well known that the solutions to the
Hamiltonian are of the form \cite{capri}
\begin{equation}
\Psi^L_n(x,y) = \exp{(i k_x x)}\psi_n[(y-y_0)/r)],
\label{landaupsi.eq}
\end{equation}
where $y_0= \hbar k_x/eB_z$ is the $y$ coordinate of the cyclotron
center, $r=\sqrt{\hbar/eB_z}$ is the magnetic length and $\psi_n$
($n$, an integer) are the normalized $n$th order Hermite
polynomials. The wavefunction $\Psi^L_n(x,y)$ characterizes the
$n$th discrete Landau level in the presence of a magnetic field,
with corresponding quantized energy spectrum $E_n =
\hbar\omega(n+\frac{1}{2})$. Although the choice of Landau gauge
preserves the translational symmetry of the system, the rotational
invariance is lost in Eq. \eqref{landaupsi.eq}. In describing the
circular Landau orbits of electrons which form in the presence of
$\mathbf{B}$ fields, it is more natural to use the rotationally
isotropic symmetric gauge. The use of the symmetric gauge has been applied previously to analyze other systems exhibiting rotational symmetry, e.g.\ in 2D two-electron systems \cite{goro}, and quantum dots (QDs) in 2D parabolic confinement potentials (see, for example, \cite{noboru,davies}), and QDs in radially symmetric hard-wall potentials with SOC \cite{tsit}, in the presence of magnetic fields.
Under this choice of gauge, it is convenient to define the complex variable $z=x+i y$ to represent
spatial coordinates within the 2DEG, and introduce the operators
\cite{capri}:
\begin{equation}
a^\dagger =\frac{r}{\sqrt{2} \hbar} (\Pi_x + i \Pi_y)\text{, }a = \frac{r}{\sqrt{2} \hbar} (\Pi_x - i \Pi_y).
\end{equation}
In analogy to the harmonic oscillator, the Hamiltonian can be
rewritten in terms of $a$ and $a^\dagger$, i.e., $\mathcal{H}_0 = \hbar \omega
(a^\dagger a + \frac{1}{2})$ with angular frequency $\omega =
eB_z/m_e$. The operators $a^\dagger$ and $a$ satisfy the usual
bosonic commutation relations and act as raising and lowering
operators on the system eigenfunctions, respectively. Through the
raising operator, we can generate the system eigenfunctions in any
level $n$, by starting with the ground state wavefunctions $n=0$ or
the lowest Landau level (LLL). The LLL is characterized by $a
\Psi_{n=0} (z) = 0$, whose solutions are given by the normalized
wavefunctions \cite{capri,laughlin}:
\begin{equation}
\Psi_{n=0,m}(z) = \frac{1}{\sqrt{2\pi r^2 2^m m!}} \left( \frac{z^*}{r}\right)^m \exp{\left(-\frac{|z|^2}{4r^2}\right)},
\label{nm.eq}
\end{equation}
where $^*$ denotes complex conjugation, and the quantum
number $m$ denotes the angular momentum. The degeneracy of the above
wavefunction in $m$ implies that one can construct general LLL
wavefunctions of the form
\begin{eqnarray}
\Psi_{n=0} &\propto & \left(\sum_m a_m (z^*)^m\right) \exp{\left(-\frac{|z|^2}{4r^2}\right)} \nonumber\\
&=& \text{(const)} f(z^*) \exp{\left(-\frac{|z|^2}{4r^2}\right)},
\label{generalpsi.eq}
\end{eqnarray}
where $f(z^*)$ is any arbitrary analytic function of $z^*$. The
normalized eigenfunctions for arbitrary $n$ and $m$ are given by
\begin{equation}
\Psi_{n,m} = \frac{1}{\sqrt{n!}} {a^\dagger}^n \Psi_{0,m}.
\label{eigenn.eq}
\end{equation}
\subsection{Landau levels with Rashba SOC and Zeeman coupling in the symmetric gauge}
We introduce spin into the system which in the case of 2DEGs in
heterostructures enters the Hamiltonian through the Zeeman coupling
and Rashba SOC \cite{rashba1,rashba2} terms. We assume a narrow-gap
heterostructure, in which the Rashba SOC term is the dominant
contribution, while the Dresselhaus SOC term \cite{dressel} can be
neglected. The Zeeman coupling and the Rashba SOC effects are,
respectively, described by the matrix operators $\mathcal{H}_Z = g \mu B_z
\sigma_z$, and $\mathcal{H}_R = \alpha/\hbar (\Pi_y \sigma_x-\Pi_x \sigma_y)$,
where $g$ is the Land\'{e} factor of electrons, $\mu=e \hbar/2m$ is
the Bohr magneton, $\sigma_{i=x,y,z}$ are the Pauli spin matrices
and $\alpha$ is the Rashba SOC parameter. In terms of the raising
and lowering operators, the Rashba Hamiltonian has the compact form
\begin{equation}
\mathcal{H}_R = \frac{\sqrt{2}\alpha i}{r} \left(
\begin{array}{cc}
    0 & a \\
    -a^\dagger & 0
\end{array}
\right).
\end{equation}
We solve the total Hamiltonian $\mathcal{H}=\mathcal{H}_0+\mathcal{H}_R+\mathcal{H}_Z$ for its eigenspinors,
$\mathbf{\Psi}_{n,m}(z)=\left(\Psi_{n,m}^\uparrow(z),\Psi_{n,m}^\downarrow(z)
\right)^\text{T}$, by writing the spinor components as a linear
combination of the spinless and normalized eigenfunctions given by
Eq.\ \eqref{eigenn.eq},
\begin{equation}
\mathbf{\Psi}_{N,m}(z)=\sum_{n=0}^{N}\Psi_{n,m}(z) \left( \begin{array}{c}a_n^\uparrow \\ a_n^\downarrow \end{array} \right),
\label{psiNm.eq}
\end{equation}
where $a_n^{\uparrow (\downarrow)}$ denotes the up (down) spin
coefficient of the $n$th Landau level, and we use the vector
notation $\mathbf{\Psi}$ to denote eigen\emph{spinor} solutions. Note
that in Eq.\ \eqref{psiNm.eq} the summation runs over the Landau
level index $n$ whilst the angular momentum $m$ is kept
constant \footnote{An equally valid basis are the functions
$\{\Psi_{n=n_0,m}\}_m$ where $n$ is fixed and $m$ run over the set
of positive integers, as one can indeed show that they form an
orthonormal set. However, in our formulation of raising and lowering
operators it is far simpler to work with our chosen basis.}. The
Schr\"{o}dinger equation $(\mathcal{H}-E \mathbf{I}) \mathbf{\Psi}_{N,m}=\mathbf{0}$
($\mathbf{I}$ is the 2-by-2 identity matrix) then reads
\begin{equation}
 \left(
\begin{array}{cc}
    (\hbar \omega (a^\dagger a + \frac{1}{2}) + g\mu B_z-E)\sum_{n=0}^{N} a_n^\uparrow \Psi_{n,m}+ \frac{\sqrt{2}\alpha i}{r}  a \sum_{n=0}^{N} a_n^\downarrow \Psi_{n,m}\\
    -\frac{\sqrt{2}\alpha i}{r}  a^\dagger \sum_{n=0}^{N} a_n^\uparrow \Psi_{n,m} + (\hbar \omega (a^\dagger a + \frac{1}{2}) - g\mu B_z-E)\sum_{n=0}^{N} a_n^\downarrow \Psi_{n,m}
\end{array}
\right) = \mathbf{0}.
\label{hamil.eq}
\end{equation}\\
To simplify Eq.\ \eqref{hamil.eq} we utilize the orthogonality of the Landau level wavefunctions, namely that $\langle \Psi_{n,m} | \Psi_{n',m}\rangle = \delta_{n,n'}$ for any value of $m$. Let us denote as $\mathbf{M}$ the column vector on the left hand side of Eq.\ \eqref{hamil.eq}. Now, we consider multiplying both sides of Eq.\ \eqref{hamil.eq} by the state-bra $\langle \Psi_{s,m} |$, which yields the equation $\int_{\mathbb{C}} \Psi_{s,m}^*(z) \mathbf{M} d^{\text{ }2}z = \mathbf{0}$, where the integration is performed over the entire complex space $\mathbb{C}$.
After canceling the orthogonal terms and applying the raising and lowering operators, Eq.\ \eqref{hamil.eq} is simplified as
\begin{equation}
 \left(
\begin{array}{cc}
    (\hbar \omega (s + \frac{1}{2}) + g\mu B_z-E)a_s^\uparrow + \frac{\sqrt{2}\alpha i}{r}  a_{s+1}^\downarrow \sqrt{s+1}\\
    -\frac{\sqrt{2}\alpha i}{r} a_{s-1}^\uparrow \sqrt{s} + (\hbar \omega (s+ \frac{1}{2}) - g\mu B_z-E) a_s^\downarrow
\end{array}
\right) = \mathbf{0}.
\label{hamilsimp.eq}
\end{equation}
The resulting equation is a simple system of two equations relating
the spinor components of state $s$ and its adjacent states $s\pm 1$.
Therefore, we can replace $s\rightarrow s-1$ without any loss of
generality in the top row of Eq.\ \eqref{hamilsimp.eq}, to yield a
regular eigenvalue equation
whose energy eigenvalues $E$ are
\begin{equation}
E^\pm = s\hbar \omega \pm \sqrt{\xi^2+2 s (\alpha /r)^2}
\end{equation}
where $\xi = \hbar \omega/2-g\mu B_z$. In particular, $\xi$ is the
energy of the LLL with eigenspinors
$(a_{s-1}^\uparrow,a_s^\downarrow)^\text{T} = (0,1)^\text{T}$.
Compared to the energy spectrum of pure Landau levels, the LLL
energy differs only by the Zeeman term $-g\mu B_z$ corresponding the
electron spins pointing antiparallel to the applied magnetic field.
Furthermore, in the LLL the wavefunctions do not experience any spin
splitting from the Zeeman term (since all eigenstates are spin down,
$\sigma_z = -1$) and the wavefunctions are completely independent of
the Rashba SOC in the system. In general, when $s \neq 0$, the
Zeeman and Rashba SO coupling breaks the spin degeneracy and the
wavefunctions are highly dependent on the SOC. Let us label the
spin-split states $\mathbf{\Psi}_{s,m}^\pm$, such that
$\mathcal{H}\mathbf{\Psi}_{s,m}^\pm = E_{s}^\pm \mathbf{\Psi}_{s,m}^\pm$.
The eigenspinor solutions are given by
\begin{equation}
\mathbf{\Psi}_{s,m}^\pm  =   N_s^{\pm}\left(
\begin{array}{c}
\kappa_s^{\pm 1} \Psi_{s-1,m} \\ \Psi_{s,m}
\end{array}
\right),
\label{psispinor.eq}
\end{equation}
where $\kappa_s= \frac{i \alpha \sqrt{2s}/r}{\xi+\sqrt{\xi^2+2 s (\alpha
/r)^2}}$, and $N_s^{\pm}$ are the normalization constants. Since the basis
wavefunctions $\{\Psi_{s,m}\}$ are normalized, $N_s^{\pm}$ satisfies $|N_s^{\pm}| =
1/\sqrt{|\kappa_s^{\pm 1}|^2+1}$. Once again, we find that the Landau
levels are infinitely degenerate since the choice of $m$ does not
affect the energy eigenvalue. Therefore, taking arbitrary linear
combinations in $m$ of the wavefunctions yield general solutions as
before:
\begin{equation}
\mathbf{\Psi}_{s}^\pm = N \sum_m a_m  \left(
\begin{array}{c}
\kappa_s^\pm \Psi_{s-1,m} \\ \Psi_{s,m}
\end{array}
\right),
\end{equation}
where the normalization constant is determined by the requirement
$\langle \mathbf{\Psi}_{s}^\pm | \mathbf{\Psi}_{s}^\pm \rangle = 1$, i.e.\
$|N|=\left(\sum_m |a_m|^2\right)^{-1/2}$
$\left(|\kappa_s^\pm|^2+1\right)^{-1/2}$.
\subsection{Gauge invariance}
We demonstrate gauge invariance of our solutions obtained in the
symmetric gauge with respect to the wavefunctions in the Landau
gauge. The U(1) gauge invariance of electromagnetism requires that
for a gauge transformation, $\mathbf{A} ^{\text{ }'} = \mathbf{A} + \nabla
\chi$, the electron wavefunction must undergo a corresponding
transformation,
\begin{equation}
\psi ^{'} = U\psi = \exp{\left(-\frac{ie}{\hbar c}\chi\right)} \psi,
\label{unitary.eq}
\end{equation}
in order for the Schr\"{o}dinger equation to remain invariant in form.
In other words the electrons acquire an extra phase factor due to the gauge transformation, which implies that physical observables are identical in both gauges. In going from the Landau to the symmetric gauge, the required gauge transformation is given by
\begin{equation}
\chi = \frac{B_z xy}{2}.
\label{chi.eq}
\end{equation}
For simplicity, we illustrate the
principle for only the $s=1$ eigenstates. Without any loss of generality,
we can focus on the eigenfunctions that have their cyclotron centres
at the origin of the system of coordinates, $(x_0,y_0)=\mathbf{0}$.
Under these set of conditions, the normalized eigenfunctions in the Landau gauge have form \cite{wang,tan}:
\begin{equation}
\mathbf{\Psi}^{L}(x,y) = \frac{N}{\sqrt{\sqrt{\pi} r}} \left(
\begin{array}{c} \kappa_{s=1}^{\pm}\\ \sqrt{2}y/r \end{array}
\right)\exp{\left(-\frac{y^2}{2r^2}\right)},
\end{equation}
On the other hand, considering
Eqs.\ \eqref{generalpsi.eq}, \eqref{eigenn.eq} and
\eqref{psispinor.eq}, the general $s=1$ wavefunctions
in the symmetric gauge are of the form
\begin{equation}
\mathbf{\Psi}^{\pm}(z) = N \left( \begin{array}{c} \kappa_1^{\pm1}
f(z^*) \exp{\left(-\frac{|z|^2}{4r^2}\right)}\\
(2\partial_{z^*}-z/2r^2) \left[f(z^*)
\exp{\left(-\frac{|z|^2}{4r^2}\right)} \right]\end{array} \right)
\label{psi.eq}
\end{equation}
Note that Eq. \eqref{psi.eq} is obtained after normalizing
$\Psi_0(z)$ and $a^\dagger\Psi_0(z)$, and substituting the explicit
expression for the $a^\dagger$ operator. Now, gauge invariance is
valid if the \emph{same} wavefunctions in the respective gauges are
linked via the relation of Eq. \eqref{unitary.eq}. It therefore
suffices to construct a wavefunction in the symmetric gauge---via
the analytic function $f(z^*)$---for which this holds. Consider
\begin{equation}
f(z^*)=\exp{\left(\frac{z^{*2}}{4r^2}\right)}=\sum_{k=0}^\infty \frac{1}{k!}\left(\frac{z^{*2}}{4r^2}\right)^k.
\end{equation}
Substituting this choice of $f$ into our symmetric gauge solution, we obtain after some manipulation
\begin{equation}
\mathbf{\Psi}^{\pm}(x,y) = N \left( \begin{array}{c} \kappa_1^{\pm1}
\eta^\uparrow \exp{\left(\frac{-y^2-ixy}{2r^2}\right)}\\
-\frac{2iy}{r^2}\eta^\downarrow\exp{\left(\frac{-y^2-ixy}{2r^2}\right)}
\end{array} \right)
\end{equation}
where $\eta^{\uparrow (\downarrow)}$ is the normalization
coefficient for the up (down) spin branch of the spinor. For
$\Psi_0(z)$ in the up-spin branch to be correctly normalized, we
require $\eta^\uparrow = \sqrt{\sqrt{\pi} r}$. On the other hand, in
the down-spin branch we set $\eta^\downarrow =
i\sqrt{r/\left(2\sqrt{\pi}\right)}$ to satisfy normalization for
$a^\dagger\Psi_0(z)$. This then yields for our symmetric gauge wavefunction
\begin{eqnarray}
\mathbf{\Psi}^\pm(x,y)&=& \frac{N}{\sqrt{\sqrt{\pi} r}} \left(
\begin{array}{c} \kappa_1^{\pm1} \\ \sqrt{2}y/r \end{array}
\right)\exp{\left(\frac{-y^2}{2r^2}\right)}\exp{\left(\frac{-ixy}{2r^2}\right)},
\end{eqnarray}
which is just the wavefunction in the Landau gauge multiplied by the gauge transformation phase factor:
\begin{equation}
\mathbf{\Psi}^\pm(x,y) = \mathbf{\Psi}^L(x,y)\exp{\left(\frac{-ie\chi}{\hbar}\right)}.
\end{equation}
\section{Numerical simulations of local spin polarization}
    \begin{figure}[!ht]
    \centering
    \begin{tabular}{c}
    \resizebox{0.6 \columnwidth}{!}{
    \includegraphics{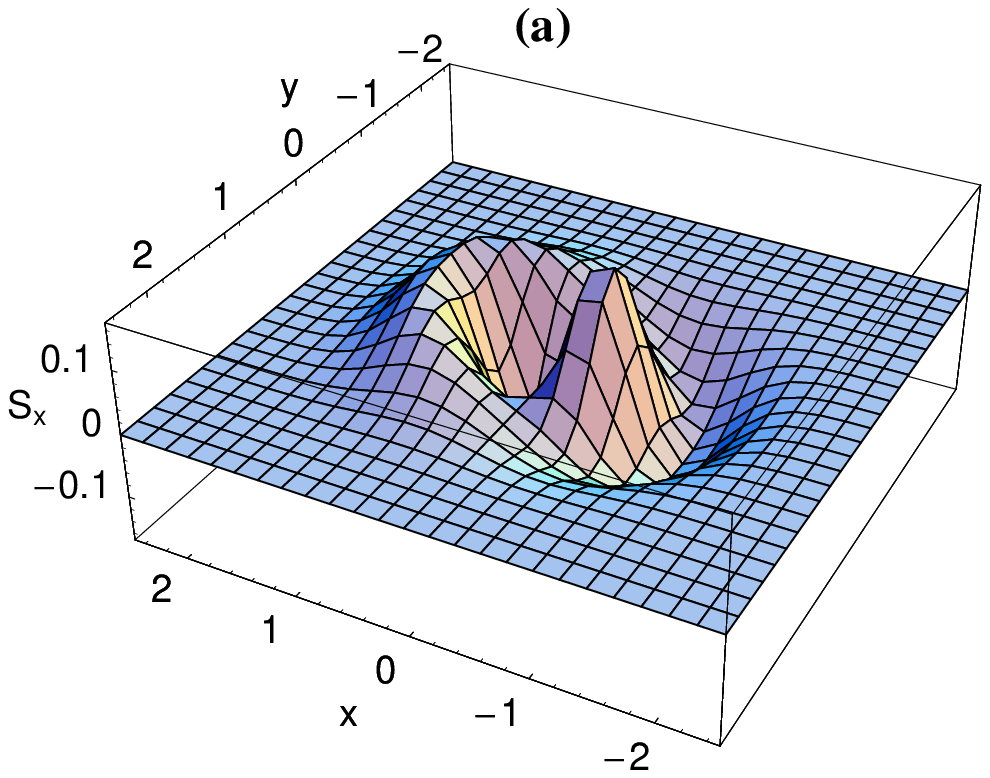}} \\
    \resizebox{0.6 \columnwidth}{!}{
    \includegraphics{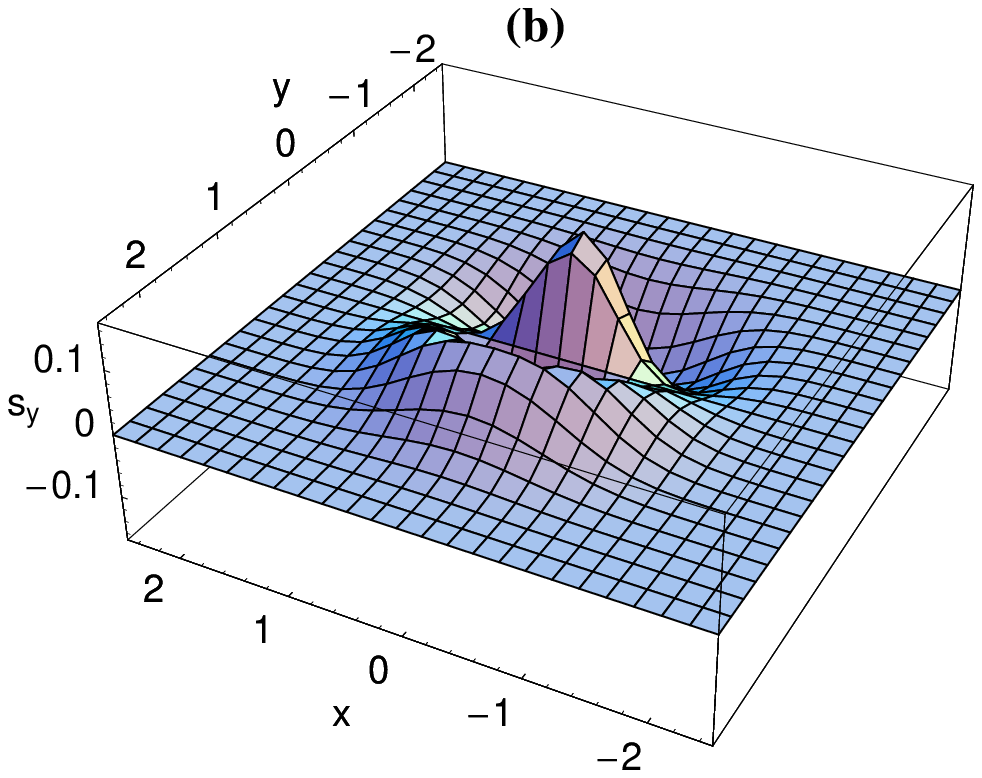}} \\
    \resizebox{0.6 \columnwidth}{!}{
    \includegraphics{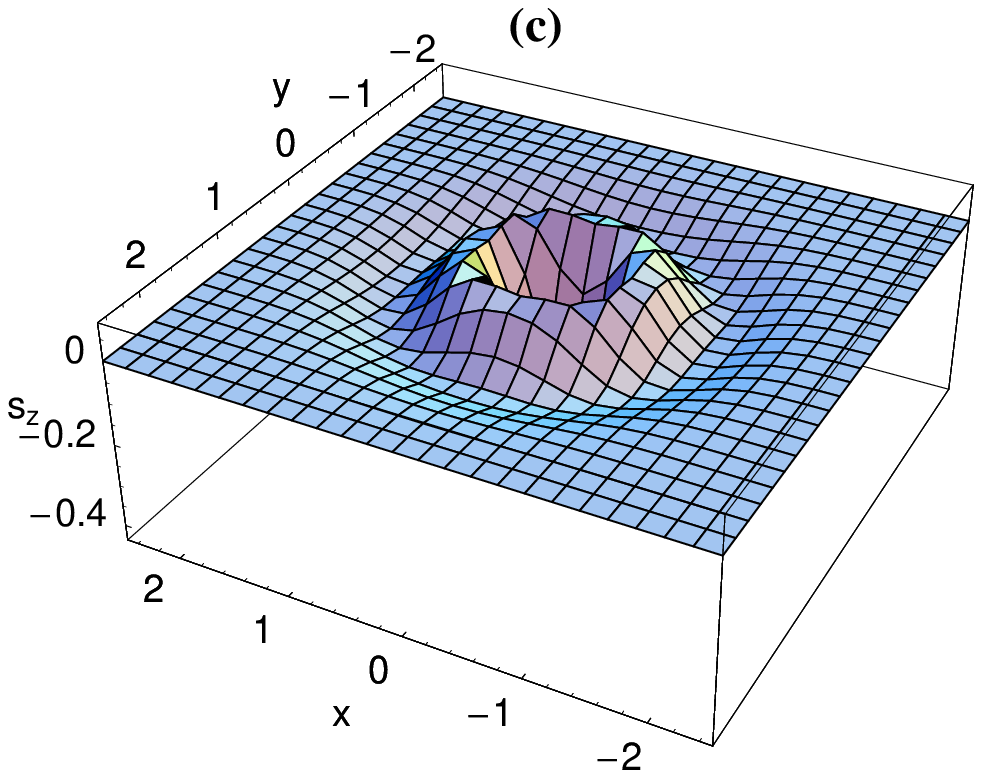}}
    \end{tabular}
    \caption{(color online) Local spatial distribution of spin polarization of $(s=1;m=1;+)$ level obtained in the symmetric gauge, $\langle \mathbf{\Psi}_{s=1,m=1}^+ | \sigma_i | \mathbf{\Psi}_{s=1,m=1}^+ \rangle$. The spatial coordinates are in arbitrary units. The $x$ and $y$ spin components, shown in (a) and (b) respectively, arise directly from the Rashba spin-orbit coupling. In (c) the spin polarization along the $z$ direction due to the Zeeman coupling is shown.}
        \label{spatialspin.fig}
    \end{figure}
We present some numerical results based on the eigenspinors we
derived for the symmetric gauge. In Fig.\ \ref{spatialspin.fig} we
plot the local, spatial spin polarization $\langle
\mathbf{\Psi}_{s,m}^\pm | \sigma_i | \mathbf{\Psi}_{s,m}^\pm \rangle$ for
the $(s=1; m=1;+)$ level in the symmetric gauge in the 2DEG plane.
First of all, we notice that the spin distributions are circular in
nature, reflecting the spatial probability density distribution of
the Landau orbits. Of the $x,y,z$-spin components, the most
interesting are the in-plane $x$ and $y$ components [Figs.\
\ref{spatialspin.fig}(a) and (b) respectively] as these components
arise from the Rashba SOC. For the Rashba Hamiltonian
$\mathcal{H}_R$, the effective magnetic field
$\mathbf{\Omega}(\mathbf{k})$ is oriented in the plane of the 2DEG, and is
orthogonal to the in-plane momentum $\mathbf{k}_\parallel$, i.e.\
$\mathbf{\Omega}(\mathbf{k}) \cdot \mathbf{k}_\parallel = 0$. The spin
alignment along $\mathbf{\Omega}(\mathbf{k})$ can be seen in Fig.\
\ref{spatialspin.fig}, if one imagines an electron moving in a
circular orbit around the origin with a tangential velocity of
$\mathbf{v}=\hbar\mathbf{k}_\parallel/m_e$. Thus, the results of our
quantum mechanical analysis are in general agreement with the
classical picture. The $z$-component of the spin, on the other hand,
shown in Fig.\ \ref{spatialspin.fig}(c) is uniform along the orbit,
as it arises from the $\mathbf{k}$-independent Zeeman coupling. For the
opposite eigenstate $(s=1;m=1;-)$, the values of the $x$ and
$y$-spin components have opposite sign. The $z$-components, however,
are not related by any simple transformation, although the general
shape of the spatial distributions is the same. Increasing $m$, one
observes an increase in the radius of the circular distributions.
Finally, the Landau level index $s$ defines the maximum number of
concentric circular orbits in the electron probability distribution.
Therefore, the electron states in the higher Landau levels are
characterized by a larger number of ``ripples'' in their spin
texture. In practice, this spatial modulation of the spin
polarization can be characterized by means of quantum point contacts
(QPC) \cite{rokhinson}, as the spatial resolution of this technique
($\sim 100$ nm) is comparable to the Landau orbital radii. In particular, one could conceive a magnetic-focusing arrangement whereby two QPCs are separated in space by twice the cyclotron radius. The first QPC behaves as an electron source, whilst the other serves as the collector. Under the influence of the magnetic field, electrons from the source follow a semi-circular trajectory in the 2DEG with a cyclotron radius, $r_c = \hbar k_F/eB$ ($k_F$ is the Fermi wavevector of electrons in the source QPC) and are collected by the detector. This technique has been used previously to image the trajectory of cyclotron orbits in 2DEG systems in the presence of a vertical magnetic field \cite{crook,aidala}. Additionally, we have shown that the momentum-dependent SOC field manifests itself as a spatially non-uniform in-plane spin-polarization along the electron orbits. This spatial variation may be detected experimentally through the use of magneto-optical techniques such as polarized absorption spectroscopy \cite{aifer}, magneto-optical Kerr rotation \cite{kikk} or magneto-reflectivity measurements \cite{chug}. A probe-based QPC technique
could also be used to tap the spin-current from the system locally
(assuming that it does not introduce significant local perturbations
to the system). By placing the probe at optimal positions
corresponding to the peak polarization values, we could conceivably
draw a highly spin-polarized current, thus implementing an efficient
spin filtering scheme.
\\
In summary, we studied the spatial spin polarization texture of Landau levels in the presence of Rashba SOC. To do so we solved the wavefunctions for the system in the symmetric gauge, demonstrating the gauge invariance of our solutions with previously known solutions in the Landau gauge. The two-dimensional analysis of the spatial spin dispersion may be important for several reasons (i) the momentum-dependent spin-orbit coupling effect is clearly seen in the Landau orbits (see Fig.\ \ref{spatialspin.fig}), and this unifies the quantum mechanical and classical pictures, (ii) the theoretically predicted spatial spin distribution may readily be verified experimentally using standard magnetic-focusing techniques, and (iii) it may find useful applications in spintronics, such as efficient spin filtering devices and sources of spin polarized current.
\section*{Acknowledgments}
The authors would like to thank the Agency for Science, Technology
and Research (A*STAR) of Singapore, the National University of
Singapore (NUS) Grant No.\ R-398-000-047-123 and the NUS Research
Scholarship for financially supporting their work.


\section*{References}
    \begin{thebibliography}{9}
     \bibitem{sa} Wolf S A, Awschalom D D, Buhrman R A, Daughton J M, von Molnar S, Roukes M L, Chtchelkanova A Y and Treger D M 2001 {\it Science} {\bf294} 1488
    \bibitem{zutic} \v{Z}uti\'{c} I, Fabian J and Sarma S D 2004 {\it Rev. Mod. Phys.} {\bf 76} 323
        \bibitem{awsch} Awschalom D D, Loss D and Samarth N (ed) 2002 {\it Semiconductor Spintronics and Quantum Computing} (Berlin: Springer)
        \bibitem{nitta} Nitta J, Akazaki T, Takayanagi H and Enoki T 1997 {\it Phys. Rev. Lett.} {\bf 78} 1335
     \bibitem{rashba1} Rashba E I 1960 {\it Fiz. Tverd. Tela (Leningrad)} {\bf 2} 1224 [1960 {\it Sov. Phys. Solid State} {\bf 2} 1109]
    \bibitem{rashba2} Bychkov Y A and Rashba E I 1984 {\it J. Phys. C} {\bf 17} 6039
    \bibitem{datta} Datta S and Das B 1990 {\it Appl. Phys. Lett.} {\bf 56} 665
    \bibitem{mireles} Mireles F and Kirczenow G 2001 {\it Phys. Rev. B} {\bf 64} 024426
    \bibitem{fujita} Fujita T, Jalil M B A and Tan S G 2008 {\it J. Phys.: Condens. Matter} {\bf 20} 115206
    \bibitem{shen} Shen S -Q, Bao Y -J, Ma M, Xie X C, and Zhang F C 2005 {\it Phys. Rev. B} {\bf71} 155316
    \bibitem{shun} Shen S -Q, Ma M, Xie X C, and Zhang F C 2004 {\it Phys. Rev. Lett.} {\bf 92} 256603
    \bibitem{zhigang} Wang Z and Zhang P 2007 {\it Phys. Rev. B} {\bf 75} 233306
\bibitem{yang} Yang C H, Xu W and Tang C S 2007 {\it Phys. Rev. B} {\bf 76} 155301
    \bibitem{wang} Wang X F and Vasilopoulos P 2003 {\it Phys. Rev. B} {\bf 67} 085313
\bibitem{demikhovskii} Demikhovskii V Ya and Perov A A 2007 {\it Phys. Rev. B} {\bf 75} 205307
\bibitem{goro} Goroshchenko S Ya and Ukrainskii I I 1988 {\it Phys. Stat. Sol. (b)} {\bf 145} 187
\bibitem{noboru} Miura N 2008 {\it Physics of Semiconductors in High Magnetic Fields} (Oxford University Press)
\bibitem{davies} Davies J H 1998 {\it The Physics of Low-dimensional Semiconductors: An Introduction} (Cambridge University Press)
\bibitem{tsit} Tsitsishvili E, Lozano G S and Gogolin A O, preprint: cond-mat/0310024 (October 2003)


     \bibitem{capri} Capri A Z 1985 {\it Nonrelativistic Quantum Mechanics} (California: Benjammin/Cummings)
     \bibitem{laughlin} Laughlin R B 1983 {\it Phys. Rev. Lett.} {\bf 50} 1395
    \bibitem{dressel} Dresselhaus G 1955 {\it Phys. Rev.} {\bf 100} 580
    \bibitem{tan} Tan S G, Jalil M B A, Teo K L and Liew T 2005 {\it J. Appl. Phys.} {\bf 97} 10A716
    \bibitem{rokhinson} Rokhinson L P, Larkina V, Lyanda-Geller Y B, Pfeiffer L N and West K W 2004 {\it Phys. Rev. Lett.} {\bf 93} 146601
    \bibitem{crook} Crook R, Smith C G, Simmons M Y and Ritchie D A 2000 {\it Phys. Rev. B} {\bf 62} 5174
    \bibitem{aidala} Aidala K E, Parrott R E, Kramer T, Heller E J, Westervelt R M, Hanson M P and Gossard A C 2007 {\it Nature Phys.} {\bf 3} 464
    \bibitem{aifer} Aifer E H, Goldberg B B, Broido D A 1996 {\it Phys. Rev. Lett.} {\bf 76} 680
    \bibitem{kikk} Kikkawa J M, Smorchkova I P, Samarth N, Awschalom D D 1997 {\it Science} {\bf 277} 1284
    \bibitem{chug} Chughtai R, Zhitomirsky V, Nicholas R J and Henini M, pre-print: cond-mat/0111492 (November 2001)
    \end{thebibliography}
    \end{document}